\documentclass[12pt]{article}
\usepackage{times}

\usepackage{graphicx}
\usepackage{fullpage,amsmath,amssymb,latexsym}
\usepackage{multirow}
\usepackage{rotating} 
\usepackage{amsmath}
\usepackage{url}
\usepackage[numbers]{natbib}
\begin{document}
\title{Saturn's fast spin determined from its gravitational field and oblateness}

%% Notice placement of commas and superscripts and use of &
%% in the author list

\author{Ravit Helled$^1$, Eli Galanti$^2$ \& Yohai Kaspi$^{2}$\\
{\small 
$^1$Department of Geosciences, Raymond \& Beverly Sackler Faculty of Exact Sciences} \\
{\small Tel Aviv University, Tel Aviv, Israel}\\
{\small$^2$ Department of Earth and Planetary Sciences} \\
{\small Weizmann Institute of Science, Rehovot, Israel}
}

\date{}
\maketitle

\begin{abstract}
The alignment of SaturnÕs magnetic pole with its rotation axis precludes the use of magnetic field measurements to determine its rotation period \cite{Sternborg2010}. The period was previously determined from radio measurements by the Voyager spacecraft to be 10h~39m~22.4s \cite{Smith1982}. When the Cassini spacecraft measured a period of 10h~47m~6s, which was additionally found to change between sequential measurements \cite{Gurnett2005,Gurnett2007,Giampieri2006}, it became clear that the radio period could not be used to determine the bulk planetary rotation period. Estimates based upon Saturn's measured wind fields have increased the uncertainty even more, giving numbers smaller than the Voyager rotation period, and at present Saturn's rotation period is thought to be between 10h 32m and 10h 47m, which is unsatisfactory for such a fundamental property. Here we report a period of 10h~32m~45s~$\pm$~46s, based upon an optimization approach using SaturnÕs measured gravitational field and limits on the observed shape and possible internal density profiles. Moreover, even when solely using the constraints from its gravitational field, the rotation period can be inferred with a precision of several minutes. To validate our method, we applied the same procedure to Jupiter and correctly recovered its well-known rotation period.
\end{abstract}

Previous theoretical attempts to infer SaturnÕs rotation period have relied on wind observations derived from cloud tracking at the observed cloud level \cite{Sanchez-Lavega2000}. One theoretical approach was based on minimizing the 100 mbar dynamical heights \cite{Anderson2007} with respect to SaturnÕs measured shape8, while a second approach was based on analyzing the potential vorticity in SaturnÕs atmosphere from its measured wind profile \cite{Read2009}. The derived rotation periods were found to be 10h~32m~35s~$\pm$13s, and 10h~34m~13s~$\pm$~20s, respectively. Our optimization method is based on linking the rotation period of Saturn with its observed physical properties and their uncertainties, in particular, the gravitational field. The method allows us to derive SaturnÕs rotation period for different types of constraints, and does not rely on a specific interior model, equation of state, wind properties, or other indirect measurements.

Previous theoretical investigations to infer Saturn's rotation
period have relied on wind observations derived from cloud 
tracking at the observed cloud level \cite{Sanchez-Lavega2000}. 
One theoretical 
approach was based on minimizing the 100~mbar dynamical
heights \cite{Lindal1985} with respect to Saturn's measured shape \cite{Anderson2007}, while a 
second approach was based on analyzing the
potential vorticity in Saturn's atmosphere from its measured
wind profile \cite{Read2009}. The derived rotation periods were found to be
10h~32m~35s~$\pm$13s, and 10h~34m~13s~$\pm$~20s, respectively. 
%The derived rotation periods were found to be

%Our optimization method is based 
%on linking the rotation period of Saturn with its observed physical
%properties and their uncertainties, 
%, in particular, the gravitational moments $J_{2n}$. The 
%and allows to derive Saturn's rotation period for different types of constraints.
% and without relying on a specific interior model, equation of state, or indirect measurements such as surface winds that have large variability or kilometric radiation measurements. 
%In addition, the method allows for a large range of solutions constrained by the measured
%physical properties of the planet and their uncertainties. 
%These solutions 
%are then used to determine the rotation period of Saturn and its uncertainty.
The gravitational moments and the internal density profile can
be related through the smallness parameter $m=\omega^{2}R^{3}/GM$,
where $R$ is the planet's mean radius, $M$ is its mass,
$G$ is the gravitational constant, and $\omega=2\pi/P$ is the angular
velocity associated with the rotation period $P$ \cite{Zharkov1978,Hubbard1984}. The even gravitational
moments can be expanded as a function of $m$ by $J_{2n}=\sum\limits_{k=n}^{3}m^{k}a_{2n,k}$, where $a_{2n,k}$ are coefficients that are determined by the radial density distribution (see Methods). 
The expansion can go to any order of $n$; since at present only $J_{2}$, $J_{4}$ and $J_{6}$ are
known for Saturn (and Jupiter), in this study we take
$n=1,2,3$. 

The relation for $J_{2n}$ shows that the measured gravitational moments are determined from the combination of the internal
density distribution ($a_{2n,k}$) \textit{and} rotation ($m$). Our goal is to find a solution for $m$ and $a_{2n,k}$
that minimizes the difference between the observed and calculated
$J_{2n}$ within the observed uncertainties. The
$a_{2n,k}$ can be expressed by a combination of figure functions (see Methods)
that represent a given internal density
profile, and can then be linked to the
gravitational harmonics \cite{Zharkov1978,Schubert2011,Kaspi2013c}.
However, since in this case there are only three equations and seven unknowns (six figure functions and the smallness parameter), there is no unique solution.
As a result, the solution is found by using a statistical optimization
approach.

%Each variable has a pre-defined range and their combination is used
%to determine the overall parameter space in which we search for solutions. 
%The relations between the figure functions and $m$ are determined for a continuous density distribution represented by a polynomial \cite{Kaspi2013c,Schubert2011}. 

We define an optimization function as the sum of the normalized
absolute differences between the observed gravitational moments and the calculated gravitational 
moments, given by
\begin{equation}
Y= \sum \bigg(\frac{|J_{2}-J_{2}^{\rm{obs}}|}{|\Delta J_{2}^{\rm{obs}}|} + \frac{|J_{4}-J_{4}^{\rm{obs}}|}{|\Delta J_{4}^{\rm{obs}}|} + \frac{|J_{6}-J_{6}^{\rm{obs}}|}{|\Delta J_{6}^{\rm{obs}}|}\bigg),
\end{equation}
where $J_{2n}$ are the calculated 
moments, $J_{2n}^{\rm{obs}}$ are the measured moments,
and $\Delta J_{2n}^{\rm{obs}}$ are the measurement uncertainties of the 
measured gravitational moments \cite{Jacobson2003,Jacobson2006}.
 The optimization procedure begins with an initial guess of the various parameters
being randomly spread throughout the physical bounds of each parameter.
This is repeated 2000 times to achieve statistical significance. From
these 2000 cases we compute the rotation period and its standard
deviation (see Methods). An example of the derived solutions using our optimization method is presented 
in Fig.~1. 

The entire set of solutions for Saturn are summarized 
in Fig.~2 which shows $P_{\rm{calc}}$
(dots) and its 1$\sigma$ standard deviation (blue-shading). 
We first present solutions that are completely unconstrained in
radius and density structure, and where the rotation period
(gray-shading) is allowed to vary widely. 
% between 0.25~hr and 5.5~hr
%around the Voyager radio period (i.e., between $\sim$ 5~hr and
%15~hr). 
%For reference, $P_{\rm{voy}}$ is presented by the dotted-black line. 
The fact that the calculated standard deviation is much smaller than
the allowed range (blue shading being much narrower than the gray
shading) indicates that knowledge of the gravitational moments can be
used to narrow the possible range of rotation periods. 
In addition, as the initial range of the possible rotation periods is narrower, the
derived rotation period can be determined with higher accuracy. 
For the smallest range in rotation period (left dot in Fig. 2a) we derive a rotation period of 10h~43m~10s~$\pm$~4m. 
The fact that the uncertainty in rotation period is decreased
significantly without enforcing tight constraints on the model emphasizes the strength of this method. 
Nonetheless, without any constraints on the shape the solution for the
rotation period still
has a relatively large range of solutions. In
reality, occultation measurements \cite{Lindal1985,Hubbard1997,Flasar2013} provide bounds on the
shape of the planet (radius vs.~latitude), 
and as shown 
below this allows to further constrain the rotation period \cite{Anderson2007,Helled2009a,Helled2011c}.

%Therefore, considering cases for which 

%To order to further constrain the calculated rotation period, we next also account for the measured radius o
%The more constraints one introduces, the more accurate is the determination of $P_{\rm{calc}}$.  
%%For example, another physical property
%%that can constrain the planet's rotation period is its shape \cite{Anderson2007,Helled2011GRL}.
%Therefore, we next consider cases for which $R_{\rm{calc}}$ must reproduce $R_{\rm{obs}}$ within
%a given uncertainty (Fig.~2b). 
%When $R_{\rm{calc}}$ must reproduce $R_{\rm{obs}}$, within
%a given uncertainty, the rotation period can be determined more accurately (Fig.~2b). 
The best measurement uncertainty
of Saturn's radii from radio and stellar occultation is $\sim$6~km \cite{Flasar2013} although the actual uncertainty could be larger due to the unknown contribution
of the atmospheric dynamics to the measured shape \cite{Helled2013}. We therefore
explore a range of uncertainty in mean radius between 6 and
80~km. 
The results for this case 
are shown in Fig.~2b where 
$P_{\rm{calc}}$ and its standard deviation vs.~the uncertainty in
observed radius $R_{\rm{obs}}$ are shown. 
%When the shape is constrained the possible range of solution narrows significantly. 
The standard deviation (blue-shading) of $P_{\rm{calc}}$ decreases with
decreasing uncertainty in the radius. For an
uncertainty of 6 km in Saturn's mean radius we derive a rotation period of 10h~34m~22s~$\pm$~3.5m. It is clear that 
the parameter space of possible solutions narrows when the constraint
of Saturn's measured mean radius is included. Yet, geopotential variations due to atmospheric dynamics affect the shape of the planet, and
therefore caution should be taken when considering these
measurements. By taking this hierarchal approach we are able to isolate the uncertainty given estimates of shape and internal structure separately.  
%A conservative estimate with an uncertainty in radius of
%$\sim$40~km yields a rotation period of 10h~36m~30s~$\pm$~5m, thus giving a solution closer to the
%Voyager rotation period. 
More conservative uncertainties in radius (tens of kilometers) yield longer rotation periods, thus giving solutions closer to the 
Voyager rotation period (see Fig.~2b).

%Also, it is clear thatWhen the uncertainty in radius
%is relatively large the calculated period is more consistent with
%the Voyager radio period, while when the shape has to be fit more
%exactly, the derived rotation period converges to shorter periods. 

The uncertainty in $P_{\rm{calc}}$ can be decreased even further if we also limit the range of the figure functions, i.e., the density profile (Fig.~2c). 
%The results when we also limit the figure functions are shown in Fig.~\ref{fig:Solutions-for-Saturn}c. 
Limiting the figure functions to within a range implied by interior
structure models (see Methods), 
%Similar to Fig.~\ref{fig:Solutions-for-Saturn}b we show
%$P_{\rm{calc}}$ and its 1$\sigma$ standard deviation vs.~the
%uncertainty in Saturn's mean radius. In this case, 
the derived period
is found to be 10h~32m~45s~$\pm$~46s. This rotation period is in
agreement with previous calculations that derived Saturn's rotation period by using a fit to its 
measured shape \cite{Anderson2007,Helled2009a}. 
%, and is only about two minutes shorter than the rotation period of 
%10h~34m~13s~$\pm$~20s which was derived from potential vorticity considerations \cite{Read2009}. 
The fact that the rotation period is shorter than the
Voyager rotation period also implies that 
%the westerly equatorial superrotating winds will be weaker allowing a more symmetric wind
the latitudinal wind structure is more symmetric, thus containing both easterly and westerly
jets as on Jupiter \cite{Read2009}. Although the smallest possible uncertainty in
rotation period is desirable, there is a clear advantage in not
specifying  constraints on the density profile, and keeping the method
as general as possible.

Different from Saturn, the rotation period of Jupiter is well determined
due to its tilted magnetic field. Jupiter's measured rotation period (system III) is 9h~55m~29.69s \cite{Higgins1996,Porco2003}. 
In order to verify the robustness of our results we
apply this method also for Jupiter
(Fig.~3). When only the gravitational moments are used as constraints (Fig.~3a), like for Saturn, the 
uncertainty in the calculated rotation period is much smaller than
the allowed range, and converges toward Jupiter's rotation period.
%The solutions (circles) essentially coincide with the black-dotted line that corresponds to Jupiter's measured
%period (Fig.~3a), indicating that for Jupiter, the possible range
%of rotation periods we consider does not have a large effect on the
%results, and for the entire range the observed period is
%reproduced. 
Fig.~3b 
shows the sensitivity of the derived period, when the uncertainty
in period is $\pm$~0.5~hr around the measured value, for a range of
possible mean radii. Like for Saturn, the standard deviation
of $P_{\rm{calc}}$ decreases with decreasing $\Delta R$. When the variation in $R_{\rm{obs}}$ is taken to
be 6~km, a rotation period of 9h~56m~6s~$\pm$1.5m is derived, consistent with Jupiter's measured rotation period. 
When we also add constraints on the figure functions, the derived rotation period becomes 9h~55m~57s~$\pm$~40s, 
showing that our method reproduces Jupiter's rotation period successfully.

The determination of Saturn's $J_{2n}$ is expected to
improve significantly following Cassini's end-of-mission proximal orbits. To test whether a more accurate determination of Saturn's gravitational field will allow to better constrain its rotation period, we repeat the optimization 
%for Saturn
with the expected new uncertainty on the gravitational moments
($\Delta J_{2n}\sim 10^{-9}$) \cite{Iess2013} around the currently measured values. 
%The solution for Saturn accounting for the expected improvements to
%the gravitational harmonics acquired by the {\it Cassini} proximal
The solution making no assumptions on the density profile is shown in Fig.~4a. 
%Saturn's rotation period is found to be $10hr~33m~21s \pm 2.3m$. 
Since Jupiter's gravitational field will be more tightly
determined by Juno \cite{Iess2013,Finocchiaro2010} we do a similar analysis for
Jupiter (Fig.~4b). While for Jupiter the calculated 
rotation period remains the same with the more accurate gravitational 
field, for Saturn the calculated uncertainty of  $P_{\rm{calc}}$
decreases by $\sim$15\%.
%For Jupiter, a rotation period of $9hr~56m~10s\pm 36s$ is derived. 
We therefore conclude that the future measurements by Cassini will be important for further
constraining Saturn's rotation period. \\
\par

{\small
\noindent{\bf Acknowledgements }
We thank Gerald Schubert, Morris Podolak, John Anderson, Lior Bary-Soroker, and the Juno science team for valuable discussions and suggestions.  We also thank the two anonymous referees for valuable comments that improved the paper. We acknowledge support from the Israel Space Agency under grants 3-11485 (RH) and 3-11481 (YK). }

\pagestyle{empty}
\begin{figure}[h]
\begin{centering}
\includegraphics[scale=0.7]{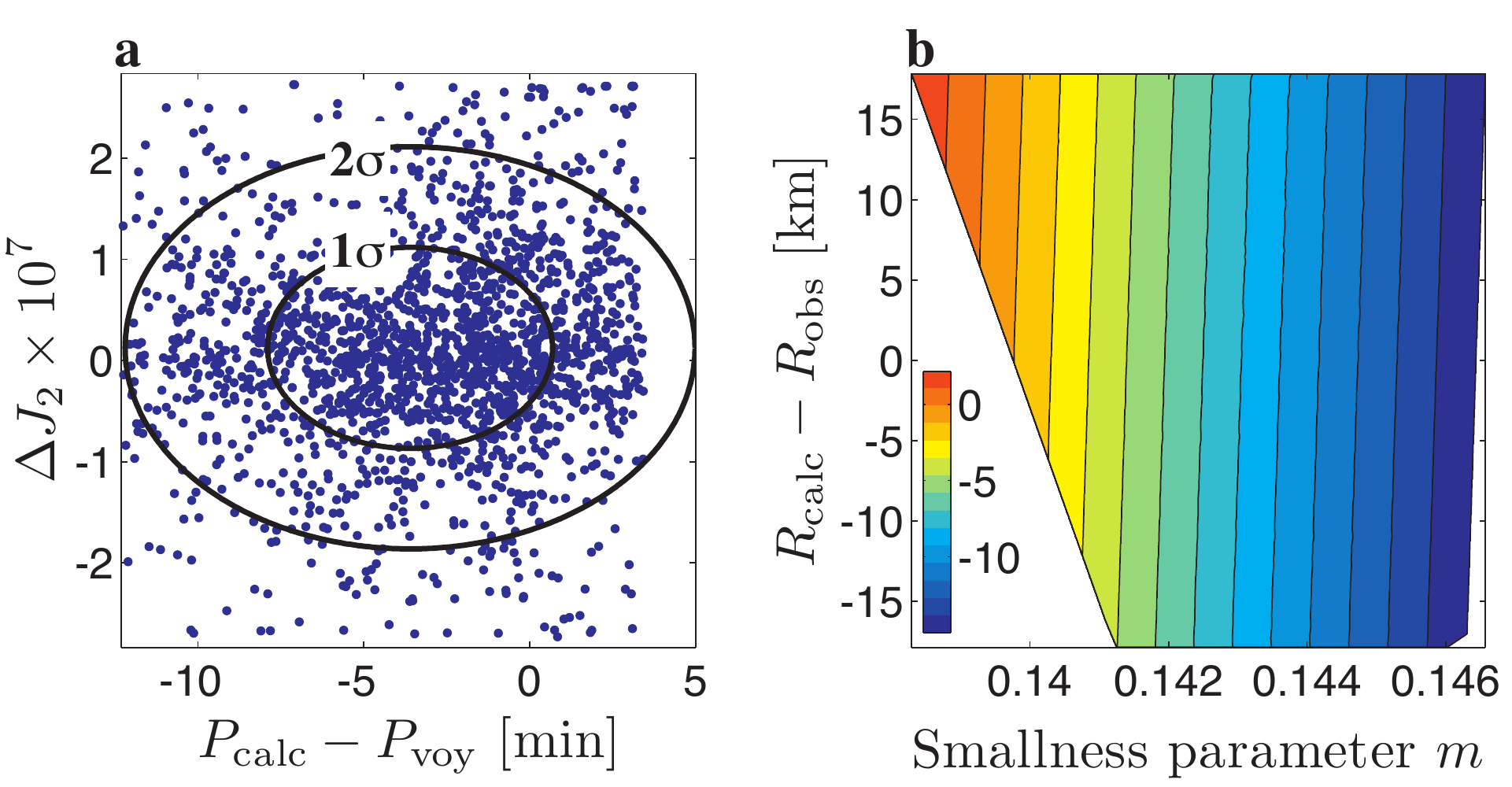}
\caption{{\bf An example of the statistical distribution of solutions for Saturn's rotation period.} 
%For this specific case, the range of rotation period is $10hr~24m - 10hr~54m$. 
For this specific case, the initial possible range of rotation
periods is taken to have 0.5~h uncertainty around the Voyager radio period (hereafter, $P_{\rm{voy}}$). 
The calculated mean radius was set to be within 20 km of Saturn's observed mean radius. 
The solution
is based on an ensemble of 2000 individual sub-cases, each of them
representing a case with specific random initial conditions within
the defined parameter space. \textbf{a.} A scatter plot of the distribution
of solutions on the plane of the calculated rotation period $P_{\rm{calc}}$
minus $P_{\rm{voy}}=$ 10h~39m~22s and $\Delta J_{2}$.
Each blue dot represents one sub-case solution. The inner and outer
black circles show the first and second standard deviations, respectively.
\textbf{b.} The distribution of the derived rotation period as a function
of smallness parameter $m$ and the calculated mean radius $R_{\rm{calc}}$
minus the observed mean radius of Saturn ($R_{\rm{obs}}=58,232$~km). 
\label{fig:A-scatter-example}}
\end{centering}
\end{figure}

\begin{figure}
\begin{centering}
\includegraphics[scale=0.7]{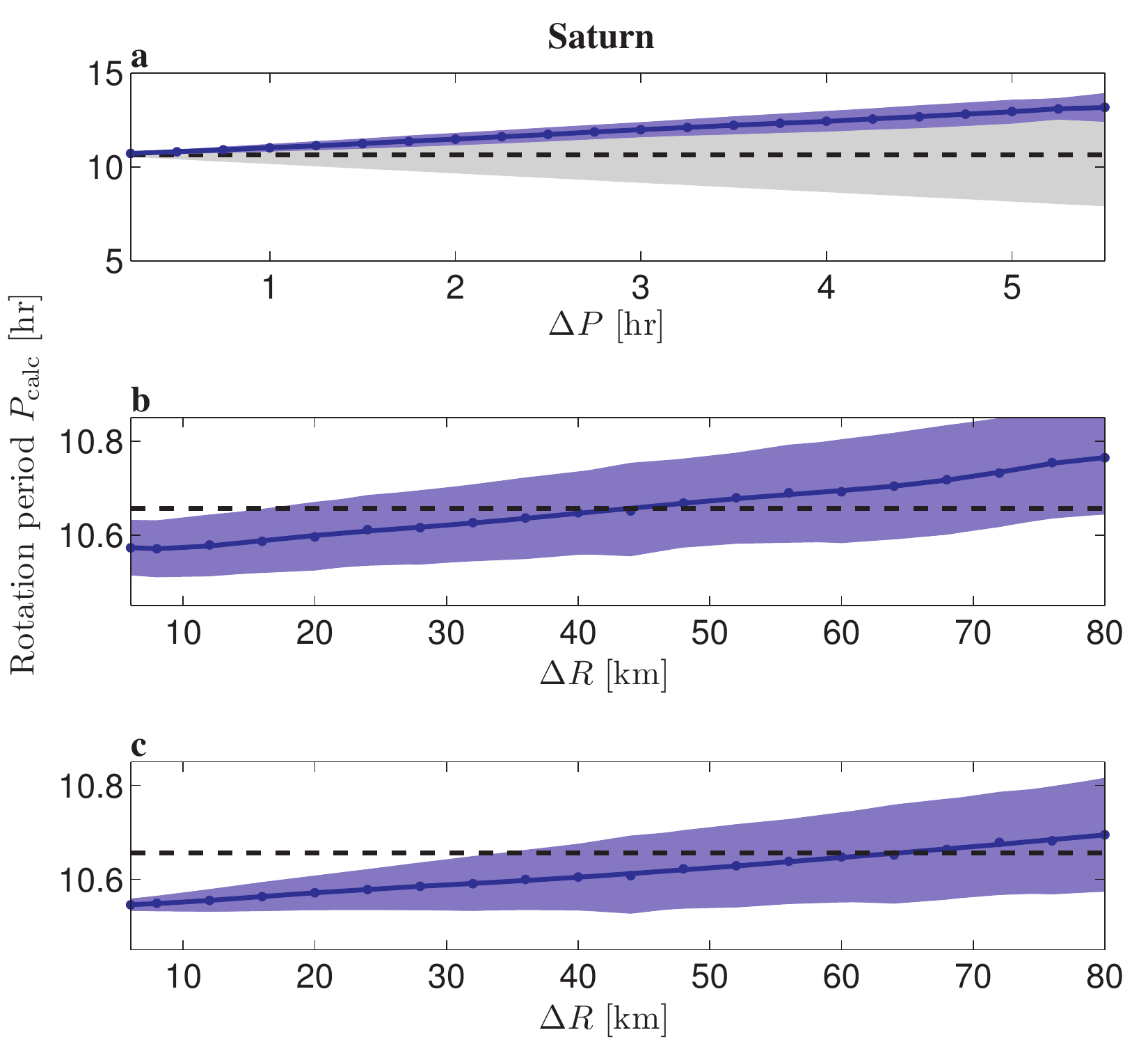}
\caption{{\bf Solutions for Saturn's rotation period.}~\textbf{ a.} The calculated period $P_{\rm{calc}}$ (blue dots) and its $1\sigma$ standard deviation
(blue shading) for a large range of cases for which the assumed possible
range in rotation period varies between 0.25~h and 5.5~h 
(gray shading) around $P_{\rm{voy}}$ (black-dashed line).
%Here Saturn's calculated mean radius is not constrained by measurements
\textbf{b.} $P_{\rm{calc}}$ and its $1\sigma$ standard deviation
(blue shading) using $\Delta P=0.5$~h vs.~the assumed uncertainty in Saturn's observed mean radius $R_{\rm{obs}}$.
\textbf{c.} Same as (b) but when the figure functions are also constrained.  
\label{fig:Solutions-for-Saturn}}
\end{centering}
\end{figure}

\begin{figure}
\begin{centering}
\includegraphics[scale=0.7]{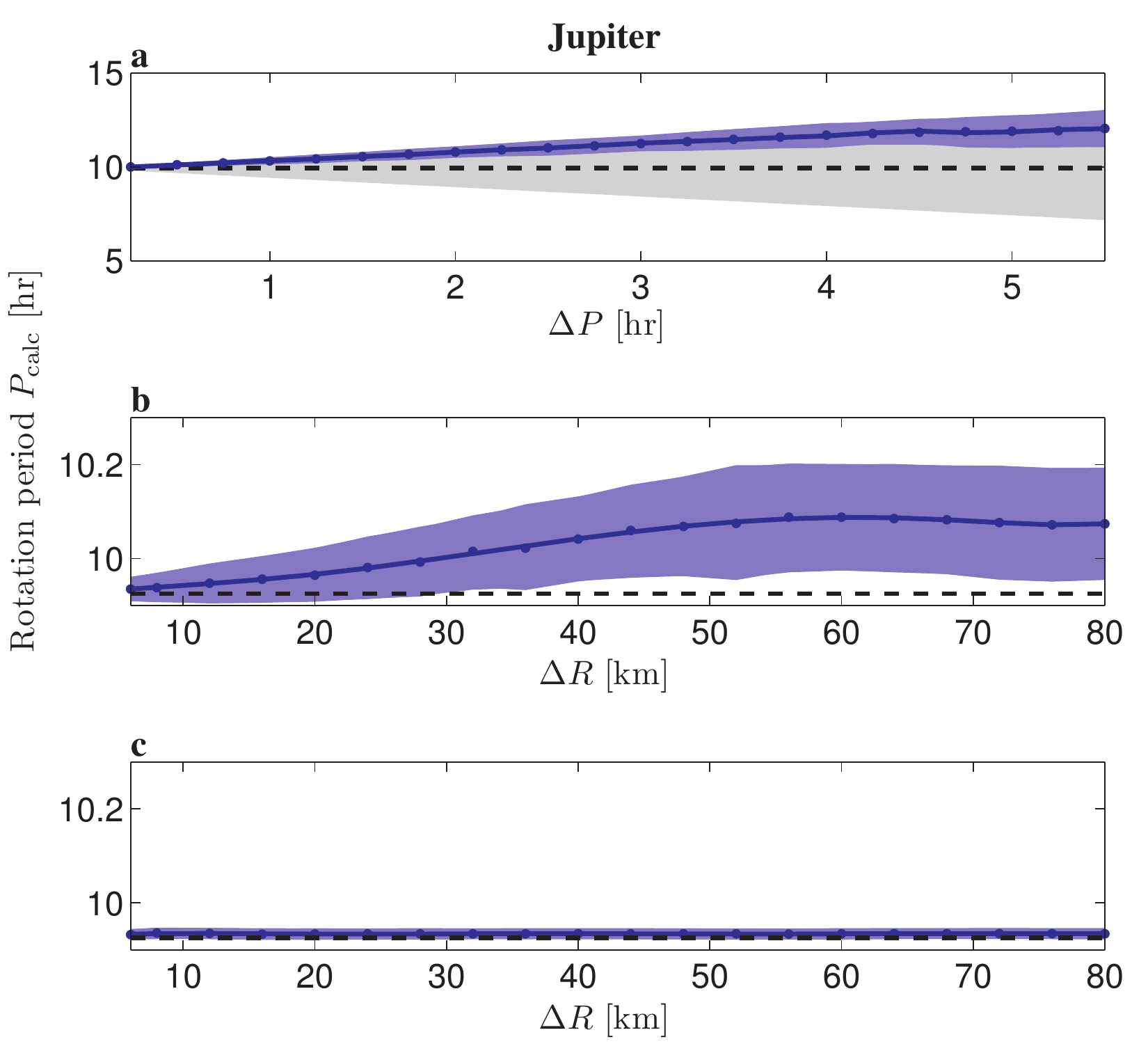}
\caption{{\bf Solutions for Jupiter's rotation period.} \textbf{ a.} The calculated period $P_{\rm{calc}}$ (blue dots) and its $1\sigma$ standard deviation
(blue shading) for a large range of cases for which the assumed possible
range in rotation period varies between 0.25~h and 5.5~h, i.e., between $\sim$ 5~h and
15~h (gray shading) around Jupiter's measured period (black-dashed line).
%Here Saturn's calculated mean radius is not constrained by measurements
\textbf{b.} $P_{\rm{calc}}$ and its $1\sigma$ standard deviation
(blue shading) using $\Delta P=0.5$~h vs.~the uncertainty in the assumed uncertainty in Jupiter's observed mean radius $R_{\rm{obs}}$.
\textbf{c.} Same as (b) but when the figure functions are also constrained. \textbf{\label{fig:Same-as-Fig.}}}
\end{centering}
\end{figure}

\begin{figure}
\begin{centering}
\includegraphics[scale=0.7]{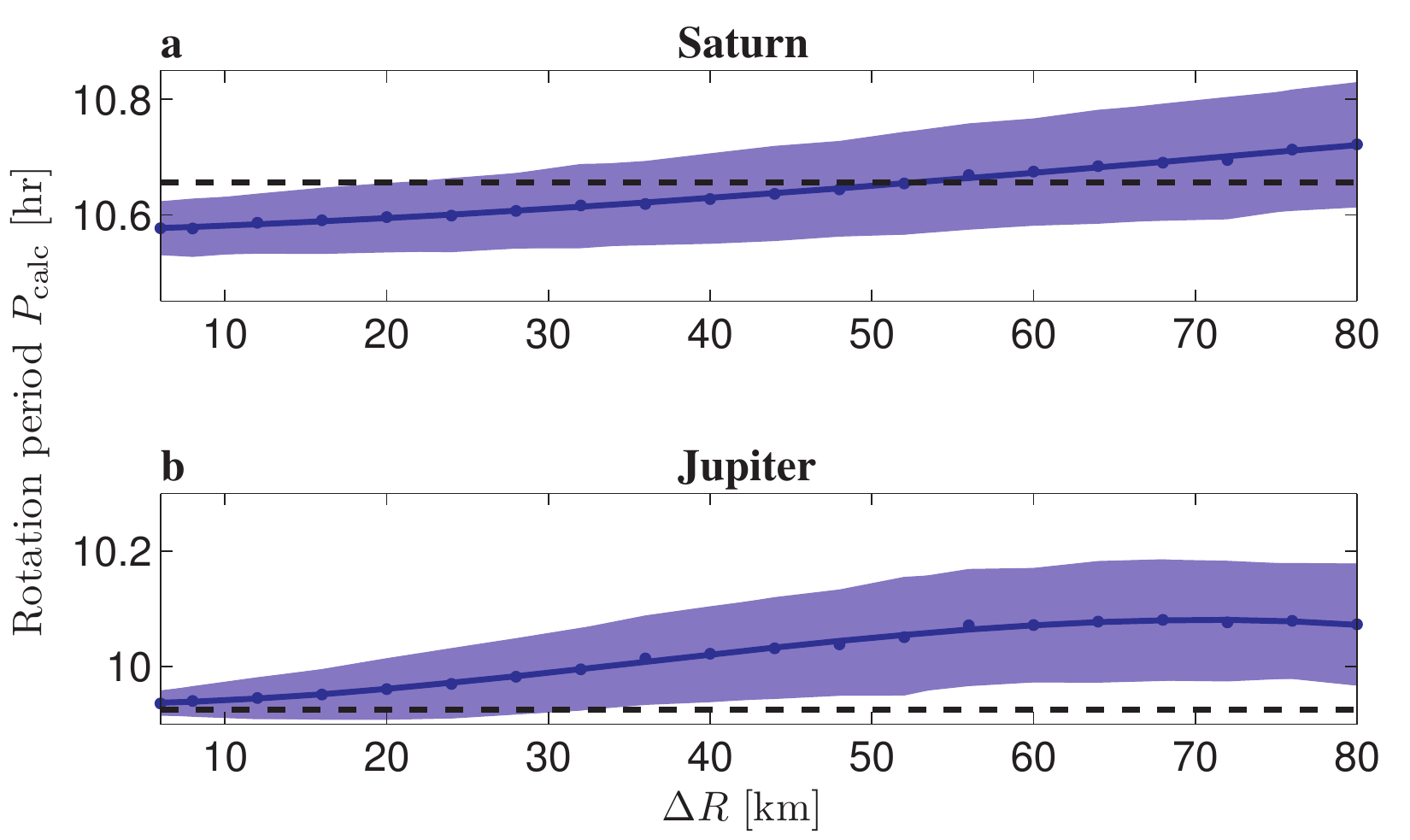}
\caption{{\bf Solutions for the rotation periods of Saturn {\bf(a)} and Jupiter {\bf(b)} when assuming improved gravity data. } 
Shown are $P_{\rm{calc}}$ and its $1\sigma$ standard deviation (blue shading) when setting assuming $\Delta J_{2n}=10^{-9}$ and $\Delta P=0.5$~h. The calculated period is given vs.~the assumed uncertainty in the observed mean radius $R_{\rm{obs}}$. 
\label{fig:new_delJ2n}}
\end{centering}
\end{figure}

\clearpage
\section{Methods}
\nocite{Sternborg2010,Smith1982,Gurnett2005,Gurnett2007,Giampieri2006,Sanchez-Lavega2000,Lindal1985,Anderson2007,Read2009,Zharkov1978,Hubbard1984,Zharkov1978,Schubert2011,Kaspi2013c,Jacobson2003,Jacobson2006,Lindal1985,Hubbard1997,Flasar2013,Anderson2007,Helled2009a,Helled2011c,Flasar2013,Helled2013,Read2009,Higgins1996,Porco2003,Iess2013,Finocchiaro2010}

\subsection{The theory of figures}
The theory of figures was first introduced by Clairaut \cite{Clairaut1743} who
derived an integro-differential equation for calculating the oblateness of a rotating 
planet in hydrostatic equilibrium with a non-uniform density profile. The method was further developed by Zharkov \& Trubitsyn \cite{Zharkov1978}
who presented a theoretical description to connect the density profile of a hydrostatic
planet with its gravitational moments $J_{2n}$, extending the theory to an arbitrary order. The basic idea of the method is
that the density profile of a rotating planet in hydrostatic equilibrium
can be derived by defining the layers as level surfaces., i.e., surfaces
of a constant potential (called the effective potential) that is set to be the sum of the gravitational
potential and the centrifugal potential \cite{Zharkov1978,Hubbard1984}
\begin{equation}
U=\frac{GM}{r}\left(1-\sum_{n=1}^{\infty}\left(\frac{a}{r}\right)^{2n}J_{2n}P_{2n}\left(\cos\theta\right)\right)+\frac{1}{2}\omega^{2}r^{2}\sin^{2}\theta, \label{U}
\end{equation}
where $r$ is the radial distance, $a$ is the equatorial radius of the geoid, \textbf{$GM$}  
is its mass multiplied by the gravitational constant, $\theta$ is
the colatitude, and $\omega$ is the angular velocity
given by $2\pi/P$, with $P$ being the rotation period. 
The internal density profile and the gravitational moments are linked
through the smallness parameter $m=\omega^{2}{R}^{3}/GM$, where
${R}$ is the mean radius of the planet. The gravitational
moments can be expanded as a function of $m$ by,
% $J_{2n}=\sum\limits_{k=n}^{3}m^{k}a_{2n,k}$, where $a_{2n,k}$ are the third order expansion coefficients in smallness parameter. 
%
%The first three even gravitational
%coefficients are given by,
\begin{subequations}
\begin{equation}
J_{2} =  ma_{2,1}+m^{2}a_{2,2}+m^{3}a_{2,3}
\end{equation}    
\begin{equation}
J_{4}  =  m^{2}a_{4,2}+m^{3}a_{4,3}
\end{equation}    
\begin{equation}
J_{6}  =  m^{3}a_{6,3}
\label{Jij}
\end{equation}
\end{subequations}
where $a_{2n,k}$ are the expansion coefficients in smallness parameter. 
As $J_2\gg J_4\gg J_6$ higher order harmonics correspond to a higher order expansion in $m$.  
The gravitational coefficients $J_{2n}$ are determined from the combination of the internal density distribution {\it and} rotation period. As a result, unless the density profile of Saturn (or any other giant planet) is perfectly known there is no simple way to derive the rotation period and vice versa. 
\par

%As discussed in  \cite{Zharkov1978,Hubbard1984} f
For the investigation of planetary figures, the equation for level surfaces can be written in the form of a spheroid that is a generalized rotating ellipsoid. 
%For an ellipsoid with an equatorial radius $a$ and flattening $f$ the equation is,
%\begin{equation}
%\frac{r^2\cos^2\theta}{a^2(1-f^2)} +\frac{r^2\sin^2\theta}{a^2} = 1.
%\end{equation}   
Then, the planetary radius $r$ at every latitude can be expressed as a function of the polar angle $\theta$ (colatitude), and the flattening parameters $f$, $k$ and $h$ by \cite{Zharkov1978,Schubert2011}:
\begin{equation}
  r \left( \theta \right) =   a \left[ 1 - f \cos^2 \theta  - \left( \frac{3}{8} f^2 + k \right) \sin^2 2 \theta
+\frac{1}{4} \left( \frac{1}{2} f^3 + h \right) \left( 1 - 5 \sin^2 \theta  \right) \sin^2 2 \theta \right],
\label{r}
\end{equation}
where $f=(a-b)/a$ is the flattening (with $b$ being the polar radius), and $k$ and $h$ are the second-order and third-order corrections, respectively  \cite{Zharkov1978}. While $f$ is strictly the flattening of the object, $k$ and $h$ represent the departure of the level-surface from a precise rotating ellipsoid to second and third order in smallness parameter, and their values are expected to be much smaller than $f$. 
\par

To third order, the three flattening parameters $f$, $k$, $h$ at the planetary surface (the effective potential surface) can be written as sum of figure functions defined by \cite{Schubert2011}, 
\begin{subequations}
\begin{equation}
f =   m F_{1} + m^2 F_{2} + m^3 F_{3}
\end{equation}    
\begin{equation}
k =   m^2 K_{2} + m^3 K_{3}
\end{equation}    
\begin{equation}
h =   m^3 H_{3}.
\label{FKH}
\end{equation}
\end{subequations}
%where $\beta$ is the normalized mean radius. 
%In this work we put no constraints on the figure functions and they are allowed to change between -1 and 1 (i.e., no constraints on the internal density profile). However, as described above the figure functions $f, k, h$ must stay positive. 

Finally, using the relation between the first three even gravitational coefficients and the figure functions for a density profile that is represented by a 6th order polynomial \cite{Anderson2007,Schubert2011,Kaspi2013c} and by applying the theory of figures as a set of differential equations, the gravitational coefficients and the figure functions can be related as power series in the small rotational parameter $m$ (see equation 72 in Schubert et al. 2011 \cite{Schubert2011}). Since only $J_{2}$, $J_{4}$ and $J_{6}$ are currently known for Saturn (and Jupiter) we expand only up to third order in $m$. Although higher-order harmonics are not expected to be zero, the corrections will be $O(m^4)$ and therefore their contribution will be small. 
\subsection{The optimization method}
Since the flattening parameters (and figure functions) depend on the density distribution that is unknown,  we take a general approach that is designed to relate the planetary rotation period to its gravitational field without putting tight constraints on the internal structure. We therefore developed an optimization method that searches for the solutions that reproduce Saturn's measured gravitational field within the widest possible pre-defined parameter space. The figure functions ($F, H, K$) are allowed to vary over their widest possible physical range, and the smallness parameter $m$ is allowed to vary within a range that reflects the uncertainty in the rotation period $P$. A solution for these parameters is sought while meeting the requirement that Saturn's measured physical properties are reproduced. 

First, an optimization function is defined as the sum of the normalized absolute differences between the observed moments and the calculated moments and is given by:
\begin{equation}
Y= \sum \bigg(\frac{|J_{2}-J_{2}^{\rm{obs}}|}{|\Delta J_{2}^{\rm{obs}}|} + \frac{|J_{4}-J_{4}^{\rm{obs}}|}{|\Delta J_{4}^{\rm{obs}}|} + \frac{|J_{6}-J_{6}^{\rm{obs}}|}{|\Delta J_{6}^{\rm{obs}}|}\bigg),
\end{equation}
where $J_{2}$, $J_{4}$ and $J_{6}$ are the gravitational coefficients calculated using Eqs.~2a-2c, $J_{2}^{\rm{obs}}$, $J_{4}^{\rm{obs}}$ and $J_{6}^{\rm{obs}}$ are the measured gravitational moments, and  $\Delta J_{2}^{\rm{obs}}$, $\Delta J_{4}^{\rm{obs}}$ and $\Delta J_{6}^{\rm{obs}}$  are the uncertainties on the measured gravitational coefficients \cite{Jacobson2003,Jacobson2006}. Since the observations include only the first three even harmonics everything is calculated to third order, but the method can be modified to include higher order terms. The data that are used by the model are summarized in Extended Data Table 1. 
\par

Next, we minimize the optimization function $Y$ with respect to the control variables $F_{1}, F_{2}, F_{3}$, $K_{2}, K_{3}, H_{3}$ and $m$, i.e., the figure functions, and the smallness parameter.
Starting from an arbitrary initial guess for each of the seven control variables (within the predefined limits), a solution is sought such that the optimization function reaches a minimum. Several nonlinear constraints are imposed while searching for the solution. 
First, we require that the difference between each calculated and the measured gravitational coefficients must be smaller than the uncertainty of the measurement error, i.e.,
%\begin{equation}
$|J_{2}-J_{2}^{\rm{obs}}|-|\Delta J_{2}^{\rm{obs}}|<0,\:|J_{4}-J_{4}^{\rm{obs}}|-|\Delta J_{4}^{\rm{obs}}|<0$, and $\:|J_{6}-J_{6}^{\rm{obs}}|-|\Delta J_{6}^{\rm{obs}}|<0$. 
%\end{equation}mmm
Note that this requirement is additional to the minimization of $Y$ since we ask that not only the overall difference between the observed and calculated gravity moments is minimized, but that individually, each of the calculated moments stays within the uncertainty of its observed counterpart. 
\par

The parameter $f$ is the planetary flattening, as a result, $f$ must be a small positive number (for Saturn $f \sim$ 0.1). %functions  $f,k,h$, calculated using Eqs.~3a-3c must be positive and smaller than $0.5$, thus reflecting a realistic shape of the planet. 
The second and third order corrections, $k$ and $h$, are significantly smaller than $f$, but could be both positive and negative. Thus, in order to keep our calculation as general as possible we let the three flattening parameters to vary between their maximum physical values, -1 and 1. 
%We therefore, let their values to vary between$\pm 10^{-2}$, although their variation is found to be smaller compared to this range.  
% flattening $f$ to be between zero and 0.5 is should be noted that the expected ranges for these parameters in significantly smaller. 
In Extended Data Fig.~1 we show the calculated values for $f, h, k$ for Saturn for the case in which the figure functions are not constrained and $\Delta R$ is taken to be 50 km. $f$ is found to be of the order of 0.1, consistent with the measured flattening of Saturn \cite{Lindal1985,Helled2013}, while the second-order and third-order corrections are found to be of the order of $10^{-3}$. As $F_1$ is the first order expansion for $f$, $f-F_1m = O(m^2)$, meaning that for Saturn $\left|{F_1}\right|<1$. Similarly expanding recursively the other coefficients of $f$, and also for $k$ and $h$, implies that all figure functions are bound between -1 and 1. In order to keep our calculation as general as possible we let all the figure functions to vary between -1 and 1. The solution though, which must also fit the gravitational field, constrains the flattening parameters and figure functions to a much narrower range. The solution is derived by using a numerical algorithm that is designed to solve constrained nonlinear multivariable functions. We use a sequential quadratic method that formulates the above nonlinear constraints as Lagrange multipliers \cite{nocedal2006}. The optimization is completed once the tolerance values for the function ($10^{-3}$) and the constraints ($10^{-12}$) are met.

A single optimization would be sufficient if the problem was well defined. In such a case, there would have been a unique solution that is independent of the initial guess. However, since in our case there are only three equations (Eqs.~2a-2c) and seven unknowns (six figure functions and the smallness parameter), the problem is inherently ill-defined and therefore has no unique solution. Nevertheless, we can still reach a solution using a "statistical" approach in which we repeat the optimization process enough times to achieve a statistically stable solution. In each case, the initial guess of the various parameters is chosen randomly within the defined bounds of each parameter. A statistical significance is reached when we repeat the optimization 2000 times (verified with $10^4$ long optimizations). We can then use the solutions from the 2000 optimizations to compute the mean value and its standard deviation for each variable.  An example for a specific case is given in Fig.~2 of the Extended Data where the solutions for the gravitational moments (Fig.~2a-c in Extended Data) are distributed around the mean value, and the distribution of the solutions for the figure functions (Fig.~2d-j in Extended Data) has a large range. %In some cases, the solution is concentrated within a relatively narrow range ($K21,H31,F11)$, other are spread over the entire allowed range ($F31)$ or over a wide but more defined range ($K31,F21$).
From each such an experiment we eventually calculate two numbers: the mean rotation period $P_{\rm{calc}}$ and its standard deviation. 

 \subsection{Saturn's rotation period with an improved measurement of its gravitational field}
 Another objective of this research is to understand whether the improved gravity measurements of the low-order gravitational moments ($J_2,J_4,J_6$) by Cassini Solstice mission can be used to better constrain Saturn's rotation period. 
Since the data are not yet available, but are expected to be within the current uncertainty, all we can do it to estimate the rotation period and its standard deviation when the uncertainty on the gravitational harmonics ($\Delta J_{2n}$) is of the order of $10^{-9}$ while using the currently known values of $J_2$, $J_4$, and $J_6$. The result for this exercise is presented in Fig.~4 of the main paper. It is found that for Saturn this yields a 15\% improvement in the derived standard deviation of its rotation period. However, it is important to remember that the true values of the gravitational moments can be any value within the current uncertainty. We find that the order of magnitude of the standard deviation is not very sensitive to the actual value of $J_{2n}$ but is more affected by the allowed uncertainty (i.e., $\Delta J_{2n}$), we can therefore conclude that the Cassini proximal orbit measurement is useful for further constraining Saturn's rotation period.    
 
 \subsection{Accounting for the planetary shape}
 
Our optimization method can include additional constraints. Since the planetary shape could be used to constrain the rotation period \cite{Helled2009a,Helled2013}, we also run cases in which we account for Saturn's shape (see Eq.~4). 
%The planetary radius $r$ can be expressed as a function of the polar angle $\theta$ (colatitude), and the shape parameters $f$, $k$ and $h$ by  \cite{Zharkov1978,Schubert2011}:
%\begin{equation}
%  r \left( \theta \right) =   a \left[ 1 - f \cos^2 \theta  - \left( \frac{3}{8} f^2 + k \right) \sin^2 2 \theta
%+\frac{1}{4} \left( \frac{1}{2} f^3 + h \right) \left( 1 - 5 \sin^2 \theta  \right) \sin^2 2 \theta \right].
%\label{r}
%\end{equation}
Then the optimization includes the constraint that the calculated mean radius $R$ should be consistent with the mean radius that is inferred from measurements.  
Thus, the calculated mean radius ${R}_{\rm{calc}}$ should be less than a specified
 uncertainty, i.e.,
\begin{equation}
|{R}_{\rm{calc}}-{R}_{\rm{obs}}|-|\Delta {R}|<0,
\end{equation}
where ${R}_{\rm{obs}}$ is the mean radius estimated from measurements of the planetary shape, and 
 $\Delta {R}$ is the uncertainty associated with the measured radius. In the standard case we set this uncertainty to be 40 km, which is large compared to the measured uncertainty in Saturn's shape \cite{Lindal1985}. This provides a fourth equation to our optimization method and allows a considerable reduction to the rotation period uncertainty (Fig.~2b and Fig.~3b in the main paper). 

Although Saturn's measured shape (radius as a function of latitude) is well determined from occultation measurements \cite{Lindal1985,Hubbard1997}, one should note that there is a difference between the measurement uncertainty, that is estimated to be $\sim$6~km \cite{Lindal1985,Flasar2013} and the actual uncertainty that is of the order of a few tens of kilometers \cite{Helled2013}. The actual uncertainty is relatively large because the planetary measured shape is also affected by atmospheric winds which distort the hydrostatic shape. 
The equatorial region of Saturn is affected by the large equatorial winds, and indeed the dynamical heights of the equator are found to be $\sim$120~km \cite{Lindal1985,Hubbard1997,Anderson2007,Helled2013}. On the other hand, the polar region is less affected by winds, and therefore the polar radii better reflects Saturn's hydrostatic shape. There are, however, no available occultation measurements of Saturn's polar regions. In addition, Saturn's north-south asymmetry in wind structure introduces an additional uncertainty in determining its polar radius \cite{Lindal1985}. As a result, a more conservative uncertainty in Saturn's mean radius is estimated to be $\sim$40 km \cite{Helled2013}.
%Saturn's shape: Lindal Hubbard1997

Interestingly, the solution for the rotation period for the case without the shape constraint does not necessarily contain the solution when the constraint on the shape is included.  
This is caused by the fact that taking into account only the gravitational moments, for Saturn, leads to a solution with relatively long rotation periods, while the measured shape pushes to shorter rotation periods. This effect is illustrated in Extended Data Fig.~3 where the solutions for the rotation period is shown in the phase-space of the constraints for $\Delta P$ and  $\Delta R$. 
When there is no constraint on the shape, and $\Delta P$ is large (upper-right, red-region), the solution converges to a relatively long rotation period; as $\Delta P$ decreases, solutions with shorter rotation periods can be found (upper-left, blue region, see also figure 2a in main text). When the constraint on the shape is included, even when the range of the rotation period is large (blue region, bottom right), the solution converges into a short rotation period. The dashed line shows the transition between the region where the constraint on the period (above the dashed line) to the regime where the constraint on the shape is more important. For the relevant physical region we are interested in, as the constraint on the shape is increased ($\Delta{R}_{\rm{obs}}$ decreases), it becomes more important than the constraint on the rotation period and therefore it becomes more dominant leading to a shorter rotation period and outside of the range of the solution without the shape constraint.
% The constraint on the shape is stronger, and therefore the solution for the shape is less sensitive, this is demonstrated by the almost horizontal blue region of the figure. 
%This suggests that there is an inconsistency between the gravitational moments of Saturn and its shape \cite{Helled2009a}. 
This behavior does not exist in the case of Jupiter as can be seen from Figure 3a in the main text. For Jupiter, the solutions reproduce the measured rotation period even for large ranges in $\Delta P$ and $\Delta R$, and the two constraints are consistent with the same rotation period.

\subsection{Constraining the figure functions} 
We also present results for cases in which the figure functions are constrained as well (Fig.~2c and Fig.~3c in the main paper for Saturn and Jupiter, respectively). In these cases, the figure functions are limited to be within a range that is determined from realistic interior models. In order to put limits on their values, we run two limiting interior models for both Saturn and Jupiter and derive the values of the figure functions. The first case is one of a massive core for which we assume a constant-density core with a density of $\sim$1.5$\times 10^4$~kg~m$^{-3}$ reaching 20\% of the planetÕs radius. In the second case, the density is continuous and is represented by a 6th order polynomial. For this case, the first degree term of the polynomial is missing so that the derivative of the density goes to zero at the center. Another constraint sets this value to zero at the core-envelope boundary for models with cores. %Further Details on the interior models can be found in \citet{Helled2011,Kaspi2013c} and \citet{Anderson2007}. 
%Because the internal density profile is represented by a mathematical function and is not derived from an equation of state of real materials, it is possible that some
%of the derived solutionss are unphysical in the sense that they correspond to a density profile that does not reproduce the density profile inferred from state-of-the-art structure models  \cite{Guillot2005,Militzer2008,Fortney2010}. 
%However, it was found that physical structure models in fact sit within the range of the massive-core+envelope and the continuous density profile models  \cite{Kaspi2013c}. 
We then use the derived values of the figure functions for each case to limit the values of the variables $F_{1},F_{2},F_{3},K_{2},K_{3},H_{3}$.
\par

The figure functions we derive for the massive core and continuous density profile cases for Saturn and Jupiter are summarized in Extended Data Table 2. 
The two density profiles for Saturn (top) and Jupiter (bottom) are shown in Extended Data Fig.~4.  
%are $F_{11} = 0.66357,F_{21} = 0.07501, F_{31} = 0.043517, \newline
%K_{21} = 0.22691, K_{31} = -0.01039, H_{31} = 0.10074$ for the massive core case, and $F_{11} = 0.67720, F_{21} = 0.04743, F_{31} = 0.10790,K_{21}, K{}_{31} = 0.07535, H{}_{31} = 0.10074$ for the continuous density profile. 
%For Jupiter, we get $F_{11} = 0.77014, F_{21} = 0.08512, F_{31} = 0.14874, K_{21} = 0.05868, K{}_{31} = -0.00705, H{}_{31} = 0.19885$ and 
%$F_{11} = 0.76965, F_{21} = 0.08499, F_{31} = 0.15137, K_{21} = 0.05861 ,K{}_{31} = -0.01068, H{}_{31} = 0.20570$ for the massive core, and continuous density profile, respectively. 
The values of the figure functions for interior models intermediate to these extreme cases all lie within the values of the extreme cases, implying that we have taken an exclusive set of values.
In order to make sure that we account for a relatively large range of possible interior models, even within this fairly constrained case, we allow the figure function values to vary around the average value between the two models by a factor of two of the difference. The results (for Saturn) are shown in Extended Data Fig.~5. It is clear that in this case, due to the limitation on the figure functions, the parameter space of possible solutions is smaller allowing a more accurate determination of $P_{\rm{calc}}$. This range still accounts for a large variation in Saturn's density profile.  While there is an improvement in the determination of $P_{\rm{calc}}$ when the shape and figure functions are tightly constrained, there is also a clear advantage in keeping the method as general as possible. The inferred result is then not associated with a specific interior model and/or does not relay on shape measurements. Our method is therefore also useful for estimating the rotation period of giant planets with less accurate determinations of their physical properties. 
For the icy planets, Uranus and Neptune, only $J_2$ and $J_4$ are
currently known, yet a simplified version of this optimization (to second order) can be
applied and gives rotation periods within 2\% of the Voyager radio periods, allowing an independent method for estimating their rotation
periods from their gravitational moments. Furthermore, our method could also be applied for deriving the rotation periods
of exoplanets for which the gravitational moments can be estimated \cite{Carter2010,Kramm2012}.

%\noindent {\bf Author contributions}\\
%R.H. led the research. R.H. and Y.K initiated the research and wrote the paper. E.G designed the optimization approach and executed all the calculations. R.H. computed interior models and defined the parameter space for the figure functions.  
%All authors contributed to the analysis and interpretation of the results.\\
%{\bf Acknowledgements }\\
%We thank Gerald Schubert, Morris Podolak, John Anderson, Lior Bary-Soroker, and the Juno science team for valuable discussions and suggestions.  We also thank the two anonymous referees for valuable comments that improved the paper. We acknowledge support from the Israel Space Agency under grants 3-11485 (RH) and 3-11481 (YK). \\
%{\bf Correspondence} \\
%Reprints and permissions information is
%   available at npg.nature.com/reprintsandpermissions. Correspondence
%   and requests for materials should be addressed to Ravit Helled
%   ~(email: rhelled@post.tau.ac.il).\\

\clearpage
\bibliography{yohaisbib}
\bibliographystyle{natbib}

\clearpage
\begin{figure}
\begin{center}
\includegraphics[scale=0.9]{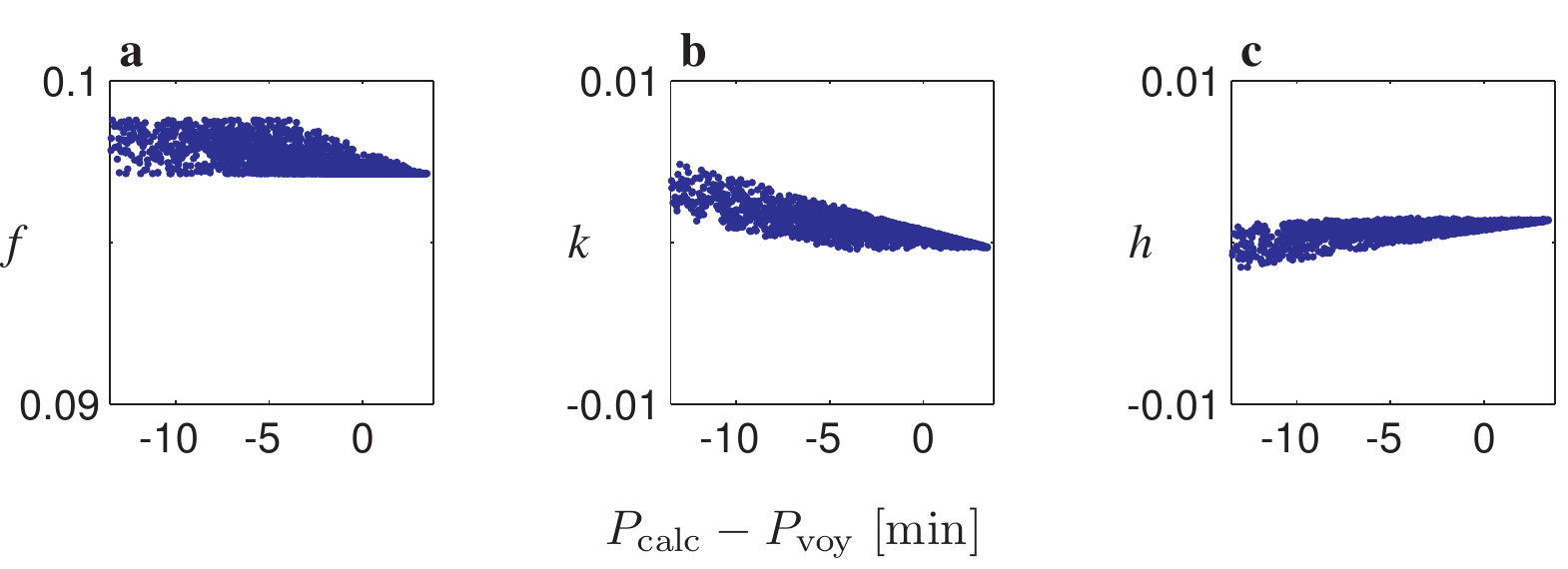}
\caption{Extended Data. {\bf The calculated flattening parameters by the model without constraining the figure functions.} Shown are $f, k, h$ for Saturn with $\Delta{R}$ = 50 km.} 
\label{fig:sup-fkh}
\end{center}
\end{figure}

\begin{figure}
\begin{center}
\includegraphics[scale=0.9]{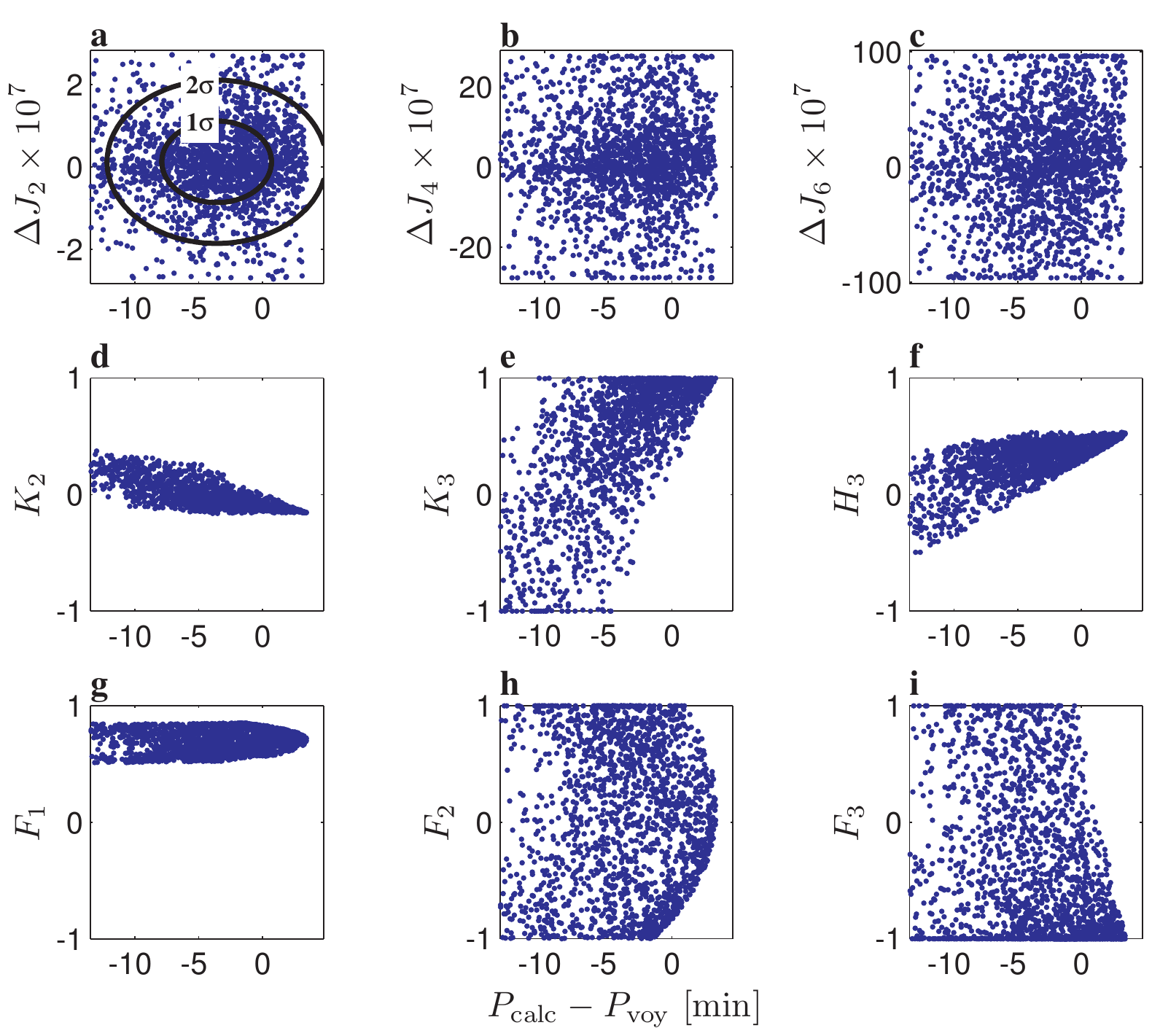}
\caption{Extended Data. {\bf An example from our statistical optimization model for deriving the rotation period.} The results are shown for a case for
which the range of rotation period is 10h~24m - 10h~54m. The solution
is based on a combination of 2000 individual sub-cases, each of them
representing a case with specific random initial conditions within
the defined parameter space. \textbf{a-c} A scatter plot (similar to Fig.~1a) of the distribution
of solutions on the plane of the calculated rotation period $P_{\rm{calc}}$
minus $P_{\rm{voy}}$ and $\Delta J_{2}, \Delta J_{4}, \Delta J_{6}$, respectively.
Each blue dot represents one sub-case converged solution. In \textbf{a} the inner and outer
black circles show the first and second standard deviations, respectively. \textbf{d-j} - the distribution
of solutions for the figure functions $K_{2}, K_{3}, H_{3}, F_{1}, F_{2}, F_{3}$, respectively.} 
\label{fig:sup-scatter-all}
\end{center}
\end{figure}

\begin{figure}
\begin{center}
\includegraphics[scale=0.7]{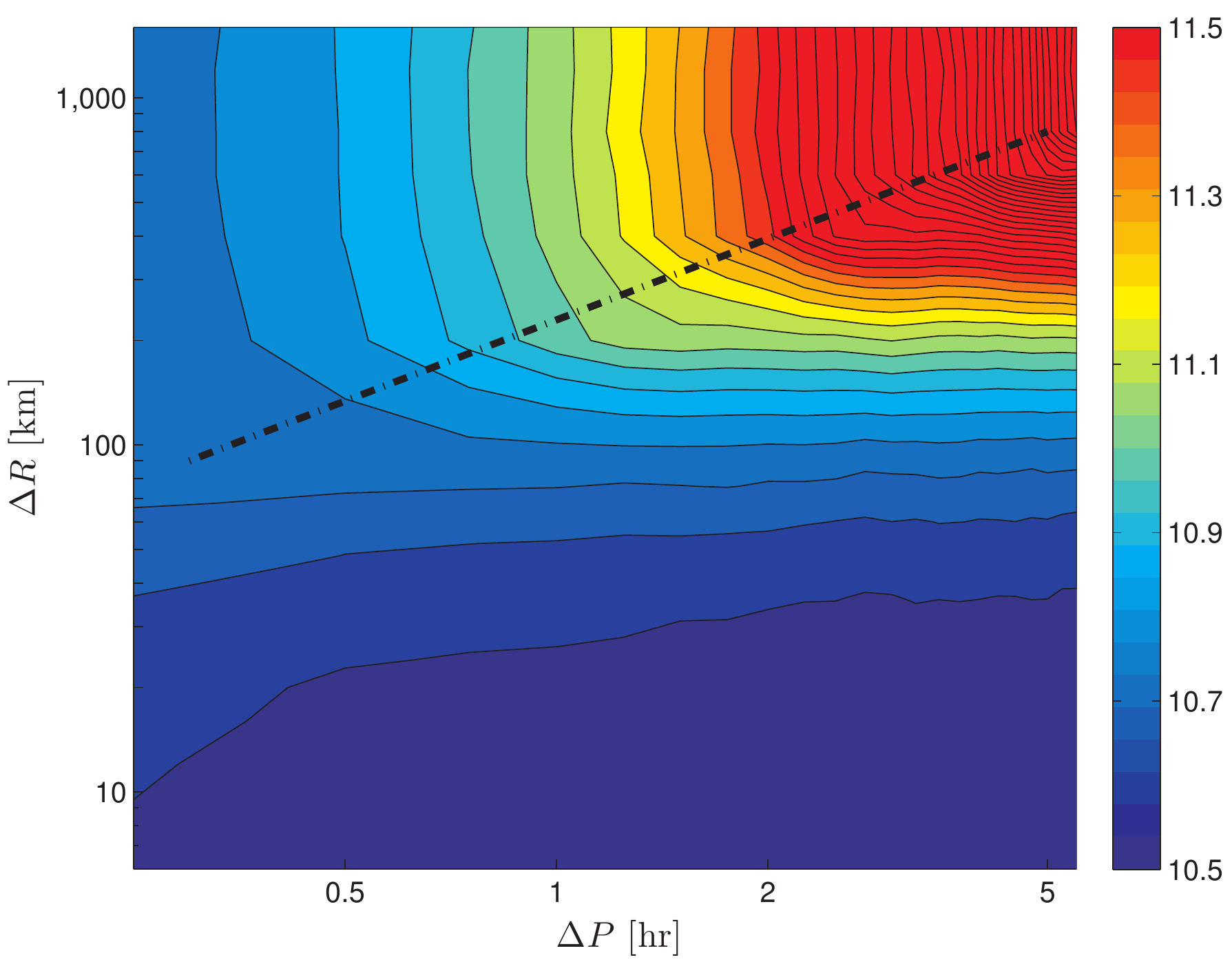}
\caption{Extended Data. {\bf Saturn's calculated rotation period vs.~the uncertainty in the assumed rotation period and radius.} Shown $P_{\rm{calc}}$ as a function of $\Delta P$ [hr] and $\Delta{R}$ [km]. The dashed line presents the transition between the region where the constraint on the rotation  period (above the dashed line) to the regime where the constraint on the shape is more dominant. }. \label{fig:sup-2D}
\end{center}
\end{figure}

\begin{figure}
\begin{center}
\includegraphics[scale=0.8]{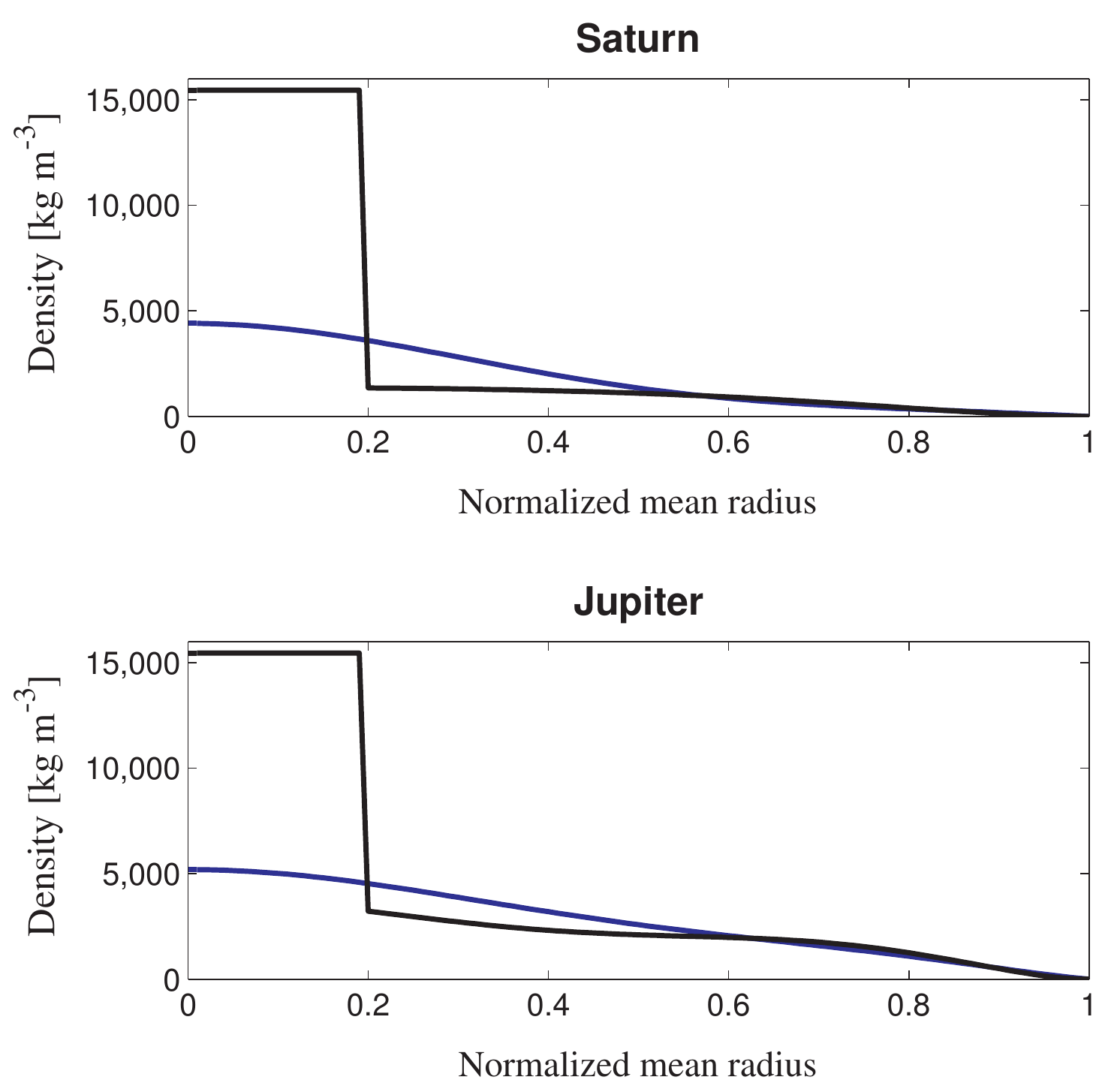}
\caption{Extended Data. {\bf Radial density profiles for two different interior models  for Saturn (top) and Jupiter (bottom).} The black curves correspond to models with very large cores while the blue curves are no-core models in which the density profile is represented by 6th order polynomials. For  the massive-core case we assume a constant core density of $\sim$1.5$\times 10^4$ kg~m$^{-3}$ reaching 20\% of the planetÕs radius. The density profiles are constrained to match the planetary mass, $J_2, J_4, J_6$,  mean radius, and the atmospheric density and its derivative at 1 bar (see details in Anderson \& Schubert, 2007; Helled et al., 2011; Kaspi et al., 2013). We then use the difference in the values of the figure functions in the two limiting cases to limit their values. } 
\label{fig:den_models}
\end{center}
\end{figure}

\begin{figure}
\begin{center}
\includegraphics[scale=0.8]{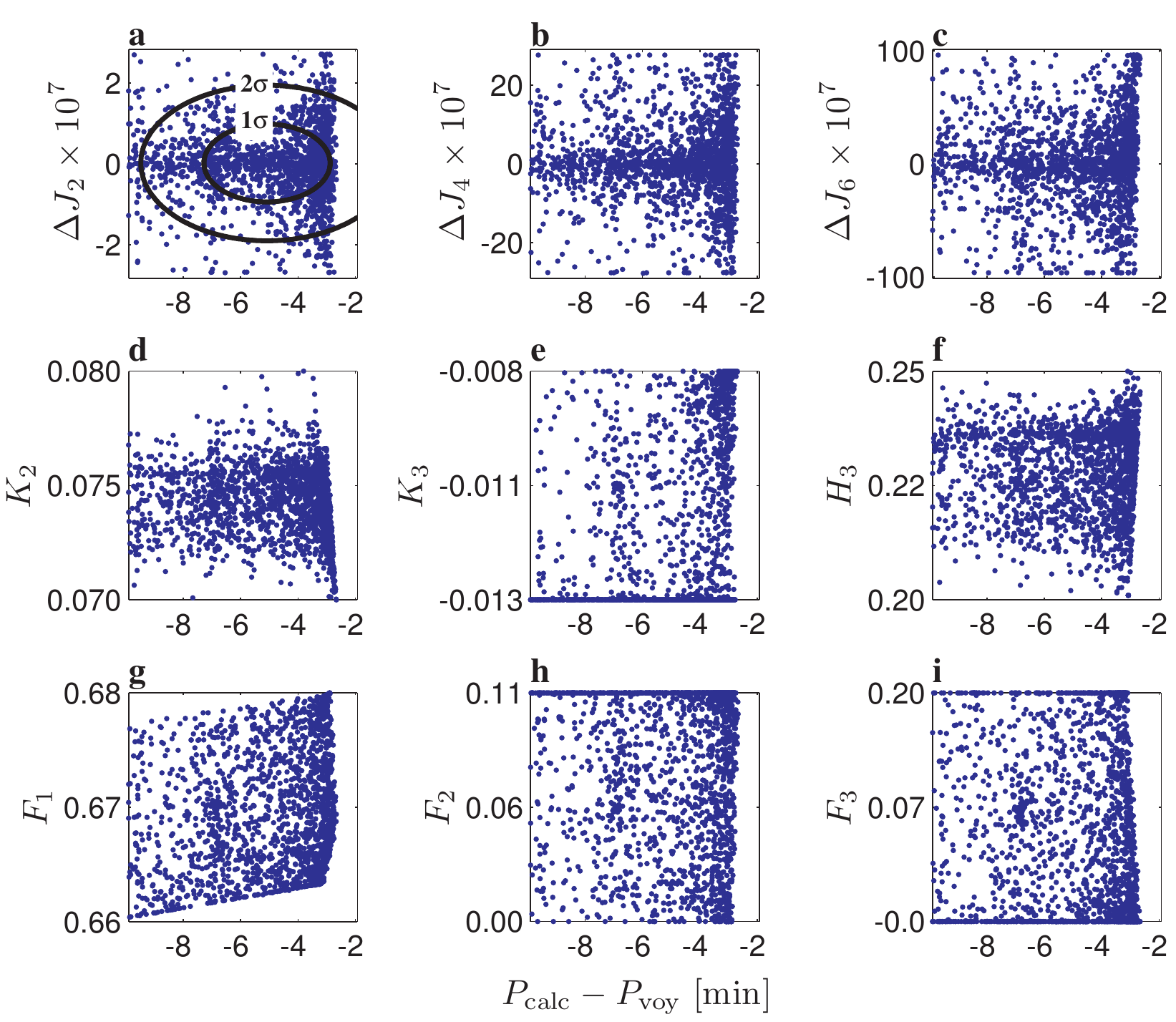}
\caption{Extended Data. {\bf The calculated flattening parameters by the model when the figure functions are limited by interior models.} 
\textbf{a-c} A scatter plot (similar to Fig.~1a) of the distribution
of solutions on the plane of the calculated rotation period $P_{\rm{calc}}$
minus $P_{{\rm voy}}$ and $\Delta J_{2}, \Delta J_{4}, \Delta J_{6}$, respectively.
Each blue dot represents one sub-case converged solution. In \textbf{a} the inner and outer
black circles show the first and second standard deviations, respectively. \textbf{d-j} - the distribution
of solutions for the figure functions $K_{2}, K_{3}, H_{3}, F_{1}, F_{2}, F_{3}$, respectively.} \label{fig:new_models}
\end{center}
\end{figure}

\clearpage
\begin{table}
\begin{center}
{\renewcommand{\arraystretch}{1}
\vskip 8pt
\begin{tabular}{l c c}
\hline
& {\bf Saturn} & {\bf Jupiter} 
\\[1pt]
\hline
Mass (10$^{24}$ kg) &  568.36 &  1,898.3
\\[1pt]
Rotation Period (System III) & 10hr 39m 24s &  9hr 55m 29s
\\[1pt]
Mean Radius (km) &  58,232  & 69,911
\\[1pt]
%Smallness Parameter $m$ & 0.14? & 0.083
%\\[1pt]
$J_2$ (10$^{-6}$) & $16,290.71\pm0.27$ & $14, 696.43\pm0.21$
\\[1pt]
$J_4$ (10$^{-6}$) & $-935.83\pm2.77$ & $-587.14\pm1.68$
\\[1pt]
$J_6$ (10$^{-6}$) & $86.14\pm9.64$ & $ 34.25\pm5.22$
\\[1pt]
\hline
\end{tabular} 
}
\caption{Extended Data. {\bf The physical properties of Saturn and Jupiter used in the analysis.} Data are taken from http://ssd.jpl.nasa.gov/?gravity\_fields\_op. 
The gravitational moments correspond to a reference equatorial radius of 60,330 km and 71,492 km for Saturn and Jupiter, respectively.}
\end{center} 
\end{table}

\newpage
\begin{table}
\begin{center}
{\renewcommand{\arraystretch}{1}
\vskip 8pt
\begin{tabular}{l c c}
\hline
& {\bf Saturn  - Massive Core } & {\bf Saturn  - No Core }
\\[1pt]
\hline
$F_{1}$  &  0.66357 &  0.67720
\\[1pt]
$F_{2}$ & 0.07501 &  0.04743
\\[1pt]
$F_{3}$ &   0.043517  & 0.10790
\\[1pt]
$K_{2}$ &  0.22691  &  0.07535
\\[1pt]
$K_{3}$ &  -0.01039 & -0.01174
\\[1pt]
$H_{3} $ & 0.10074 &  0.23489
\\[1pt]
\hline
& {\bf Jupiter  - Massive Core} & {\bf Jupiter - No Core }
\\[1pt]
\hline
$F_{1}$  &  0.77014 &  0.76965
\\[1pt]
$F_{2}$ &  0.08512 &  0.08499
\\[1pt]
$F_{3}$ &  0.14874  & 0.15137
\\[1pt]
$K_{2}$ &  0.05868 & 0.05861
\\[1pt]
$K_{3}$ &  -0.00705  & -0.01068
\\[1pt]
$H_{3} $ &  0.19885  & 0.20570
\\[1pt]
\hline
\end{tabular} 
}
\caption{Extended Data. {\bf The calculated figure functions based on interior models of Saturn and Jupiter.}
The values of the figure functions are derived for the two limiting cases of massive core and no-core (continuous density profile) for Saturn and Jupiter. }
\end{center} 
\end{table}

\end{document}